\documentclass[12pt]{amsart}

\usepackage{graphicx}
\usepackage{epsfig}
\usepackage{xcolor}
\usepackage{hyperref} %\hyperref[label]{text}
\usepackage[utf8]{inputenc}
\usepackage[russian,english]{babel}
\def\H{\mathcal H}
\def\bra{\langle}
\def\ket{\rangle}

\def\hat{\widehat}

\begin{document}
\title{Indeterminism and Classical Mechanics}
\author[Alexander Givental]{Alexander Givental}
\address{Department of Mathematics, University of California Berkeley}
\email{sumizdat@berkeley.edu} 
\dedicatory{To Richard Montgomery and Sergei Tabachnikov on their 70th birthdays}

\maketitle  
\ 
%\vspace{-10mm}
\begin{abstract} We challenge the traditional deterministic view of the classical mechanical universe by combining dynamical chaos with the distinction between physical phenomena and their mathematical models, and quantify the resulting indeterminism in several quasi-realistic examples.
\end{abstract}  

{\bf 0. Summary.} The deterministic view of the mechanical universe comes from the success of classical mechanics in capturing the dynamics of macroscopic --- celestial and domestic --- physical systems. The present note is to remind ourselves, however, that the ODE systems describing the time evolution in classical mechanical systems should not be literally identified with the actual phenomena they are supposed to model. The precisely defined phase points and parameters of the systems are merely mathematical idealizations. This observation coupled with the exponential divergence of phase trajectories shows that classical determinism is also an idealization. Not only phase trajectories can be {\em unpredictable in practice} due to the dynamical chaos, but they are {\em inherently undefined} as physically meaningful objects. Consequently, the real mechanical universe is not {\em predestined} for a particular future --- nor is it guaranteed to have a definite past. 

\medskip

{\bf 1. Sources of ambiguity.} Before accepting these conclusions, let us first go in the opposite direction and try to describe a realistic object as fully deterministic.

Let's, say, take my desk. It is a rigid body of a certain well-defined shape whose fully deterministic rotational properties are therefore described by its inertia ellipsoid which might be not too hard to compute.

Yet, under some 50-fold magnification one would find that the rectangular shapes of my desk's parts are not exactly rectangular and have hard-to-describe surfaces and edges. Still, one can argue, though the actual shape is hard to describe, the object has a certain shape, and given the fixed density of the material (let's not worry about it for the moment), the inertia ellipsoid is well-defined (though maybe hard to compute precisely) and so is the deterministic rotational behavior of this rigid body.   

Yet, it is not true that the desk is genuinely rigid. Knocking on it generates sound waves --- elastic vibrations of this continuous medium (so, let's not worry here that the vibrations are not exactly elastic due to some plasticity of every material). Still, one can imagine the PDE system describing wave propagation in this medium and conclude that the dynamics of the object is described deterministically, though this time in an infinite dimensional phase space.

Yet, at the nanoscale, the desk does not look like a continuous medium, but rather like an atomic lattice (though in reality not very regular due to numerous defects). So what? Regular or not, one can in principle model the desk as a collection of pointlike Newtonian particles interacting according to a specific interaction potential thus returning us back to a deterministic system of ODEs, though in a hugely dimensional phase space.

Thus, if we could dive indefinitely into this rabbit hole of more and more sophisticated yet fundamentally classical mechanical systems, we would argue that the behavior of the physical object is inherently deterministic, even if we cannot give any ultimate model of it.

In reality this attempt fails because at the atomic level the ingredients of the system are neither pointlike classical particles nor rigid bouncing balls, but quantum entities whose behavior is inherently random. That is, the dynamics of the quantum mechanical system is described by the deterministic Schr\"odinger equation. Yet, a {\em state} of the system whose time evolution is so described is a wave function $\psi$ whose physical significance lies in capturing expectation values $\bra \psi|A|\psi\ket$ of physical quantities $A$ in repeated identical measurements. Thus, at the bottom of the classical indeterminism we are talking about there ultimately lies the inherent randomness of quantum phenomena.

One way how this randomness manifests numerically is the Heisenberg uncertainty relation $2 \Delta q_\alpha \Delta p_\alpha \geq \hbar \approx 10^{-34} kg\cdot m^2/s$ (where $q_\alpha$ and $p_\alpha$ are corresponding components of position and momentum considered as Darboux coordinates on the phase space of the underlying classical mechanical system). Perhaps an adequate way to interpret the interplay between classical and quantum here is to assume that a point (e.g. $(p,q)=(0,0)$) in the classical phase space is represented by the Gaussian ``wave packet'' (in coordinate and momentum representation respectively):
\[ \psi \sim e^{-\sum_\alpha q^2_\alpha/2\mu\hbar} \ \text{or}\ \hat{\psi} \sim e^{-\sum_\alpha p^2_\alpha\mu/2\hbar},\ \ \mu=O(1) \ kg/s \]
This is to say that the Heisenberg uncertainty is distributed equally between positions and momenta:

$$\Delta q_\alpha \approx 10^{-17}m, \ \ \Delta p_\alpha \approx 10^{-17} kg\cdot m/s,$$
and so more precise values of these quantities make no physical sense.

Yet, there are other mechanisms of how physical quantities lose precise meaning beyond certain scale. For example, physicists say that below the Planck scale $10^{-35}m$ space itself makes no sense. The explanation goes like this. To measure distance with the precision $\Delta l$ one needs to use photons of this wavelength, i.e. of frequency $\omega = c /\Delta l$, where $c$ is the light speed,
and respectively of energy $E=\hbar \omega = \hbar c /\Delta l$. But energy is mass, and this energy concentrated in the radius $\Delta l$ would create a microscopic black hole, from which no information about the result of the measurement would escape. Anyway, regardless of the validity of the last argument, the value of the size of my desk in meters makes no sense beyond $35$ decimal places.

Thus, we have three length scales to pay attention to: the {\em atomic scale} $10^{-10}m$ below which objects lose their classical nature, the {quantum scale} $10^{-17}m$ below which Heisenberg's uncertainty of positions kicks in (though it can be improved by a greater uncertainty in momenta), and the {\em Planck scale} $10^{-35}m$ below which all lengths become physically meaningless.

Perhaps one should be more careful interpreting the last statement. For example, the coordinates of the center of mass of my desk have more significant digits. Namely, the position of each atom in it cannot be more precise, but averaging the uncertainties considered random and independent improves the precision $\sqrt{N}$ times where the number $N$ of atoms is at the Avogadro scale of $10^{24}$.     

Finally, one can think of other reasons why modeling macroscopic objects by closed classical mechanical system is inadequate. A famous example was provided by Michael Berry \cite{B} (with reference to Chirikov): in the hard-sphere model of fluid, collisions become unpredictable after several dozens of them because of the unaccounted gravitational pull by electrons located at the outskirts of the visible universe. The point there is that no particular system is fully isolated, but it leaves the loophole that the entire mechanical universe still is. Moreover, until very recently when biologists and neuroscientists started noticing quantum effects in living systems, they often maintained that electro-chemical processes in cells or brains are macroscopic enough to be fully deterministic (with implications concerning free will \cite{H}). Our goal, however, is to discredit the very idea of classical determinism, and do this on the grounds equally applicable to any isolated system, be it the entire universe or a part of it.         

\medskip

{\bf 2. Lyapunov exponents.} Physical (philosophers would say {\em ontological}) ambiguity of initial conditions would not yet lead to ontological indeterminism if the dynamics were Lyapunov stable. That is, if phase trajectories whose initial conditions vary within a small margin $\epsilon$ remain within distance $O(\epsilon)$ from each other during the entire evolution, one would conclude that, although physical trajectories in the phase space are not defined in the strict mathematical sense, they are defined with that same precision with which phase points themselves make physical sense.    

For example, in integrable systems this is true for quasi-periodic oscillations within the same invariant torus (i.e. with fixed values of action variables). On the other hand, a small change in the action variables results in a comparably small change $\Delta \omega$ of the oscillation's frequencies. Over time $t$ this leads to {\em secular perturbations} of the phase point on a scale $\sim \Delta \omega \cdot t$. One should note, however, that $\Delta \omega \sim 10^{-35}\ s^{-1}$ should be contrasted with the age of the universe $\sim 10^{18}\ s$. The conclusion is that integrable systems should be considered rather immune to ontological indeterminism: If the time needed for the inherent uncertainty of mathematical modeling to accumulate to macroscopic values exceeds the lifespan of the system, then the mathematical model has to be declared physically accurate.

Yet, {\em exponential divergence} of phase trajectories and the resulting dynamical chaos is one of typical behaviors observed in Hamiltonian systems.
Namely, a discrepancy $\epsilon$ of initial conditions can turn into the discrepancy $e^{\Lambda T} \epsilon$, where $\Lambda>0$ is a {\em Lyapunov exponent}, and $T$ is the age of the system. If a more elaborate picture is desired, one should understand that locally in a fixed coordinate system in the phase space trajectories adjacent to a given one stay within the fixed level hypersurface of the Hamilton function, but within that level and in the slice transversal to the given trajectory, the dynamics after time $t$ is described by a canonical transformation whose linearization at the central point (representing the given trajectory) has spectrum invariant under the map $\lambda \mapsto \bar{\lambda}^{-1}$. In particular, to every real $\lambda\neq \pm 1$ (or a pair of complex-conjugate eigenvalues of absolute value $|\lambda|\neq 1$) there corresponds a line (resp, plane) of exponential expansion in forward time, and another direction (resp. plane) of exponential expansion with the same rate in backward time. The eigenvalues with the maximal absolute value thus determine the typical local expansion rate (in either time direction). However, the eigenvalues $\lambda$ vary from a point to point, the respective eigenspaces don't have to be aligned, and in any case neither eigenvalues nor eigenspaces have invariant meaning (i.e. they depend on the coordinate system). So, the Lyapunov exponent $\Lambda$ mentioned above is some cumulative average characteristic of the dynamics which cannot be directly computed from the local eigenvalues, and is rather found empirically based on the long-term behavior of typical trajectories in some region of the phase space where exponential divergence does take place.

Thus, our goal is to look at some realistic dynamical systems and, assuming that $\epsilon$-discrepancies between initial conditions are (not merely irrelevant for practical purposes but) physically meaningless, find out how the time $T$ such that $e^{\Lambda T} \epsilon \sim 1$ compares to the lifespan of the system.

In what follows we give a few hand-made illustrative examples of this sort. In fact, literature examining numerically the behavior of dynamical systems, including Hamiltonian ones, and estimating their Lyapunov exponents, is enormous. Compared to that literature, our numerical simulations are extremely childish. The only reason why we nonetheless resort to such simulations instead of referring to the available literature is that usually the models in the literature are described in mathematically minimalistic and therefore {\em dimensionless} form, while for our goals we need to control the actual physical parameters of our systems.

Yet, there is at least one reason why the ultimate verdict concerning
indeterminism is predetermined before any simulations. Namely,
dynamical chaos is a real phenomenon and a serious obstruction to scientific and practical predictions of various sorts. For example, it is clear that many objects in the Solar system are monitored by NASA, that predicting their whereabouts by numerical simulation is important, and that non-integrability of the 3-body problem and any other manifestations of dynamical chaos present a serious limitation. Although we don't have direct access to astronomers' data,
it is not hard to imagine that, say, a $1,000\ km = 10^6\ m$  error in the
current position of an asteroid may result in a discrepancy at the scale of one astronomical unit $\approx 10^{11}\ m$ after, say, 10 years: $e^{\Lambda \cdot 10} 10^6 \approx 10^{11}$. The question now is how soon an error in the current position of the asteroid at the {\em Planck scale} $10^{-35}\ m$ will result in a spatial discrepancy of astronomically significant scale. The key factor here is that the time $T$ we are looking for depends on precision {\em logarithmically}:
\[ \text{If} \ 10 \Lambda  = 5 \log 10  \ \text{then} \ \Lambda T = 46 \log 10 \ \ \text{implying}\ T< 100\ \text{years} .\]
That is, if the horizon of practical unpredictability due to measurement errors is significant by the scale of human lifespan, then the horizon of ontological indeterminacy, which is only dozens of times longer (10 in this example), is certainly many orders shorter than the age of the universe.

\medskip

{\bf 3. Sinai's billiard.} The most famous examples of dynamical chaos are provided by the geodesic flows on Riemann surfaces of constant negative curvature. The Lyapunov exponents for these systems are constant and known, and so it would be very convenient to have such flows embedded into phase spaces of physically meaningful mechanical systems. Unfortunately I am not aware of such embeddings, but here is a less symmetric dynamical system which also is well-known to exhibit exponential divergence of trajectories.

Consider a square billiard table of size $L\times L$ with a reflecting disk of radius $R(<L/2)$ at the center of the square. One can think of a billiard as a closed surface glued of two sheets along their boundaries, and a billiard trajectory as a geodesic crossing from one sheet to the other at the moment of reflection at the boundary. In such terms, our billiard is a torus whose total curvature, which by the Gauss-Bonnet theorem equals $2\pi$ times the Euler characteristic of the surface, is therefore zero. Namely, four angular defects at the corners of the square contribute $+\pi/2$ each, while the compensating $-2\pi$ is distributed uniformly along the circular boundary in the middle of the square. A random trajectory has probability zero to hit a corner, but the reflections off the concave circular boundary have a dispersing effect, and this can be taken as an explanation for the overall exponential divergence of trajectories.

Another way of looking at the billiard is to replace reflections of a flat wall by the reflection of the billiard itself. This way Sinai's billiard becomes equivalent to {\em Lorentz' gas}: a pointlike particle (an electron in the original interpretation) bouncing off radius-$R$ balls (nuclei of a crystal) positioned at the vertices of an infinite square lattice with step $2L$. The genus of the corresponding two-sheeted Riemannian surface is infinite (which indicates that dynamical chaos in billiards is rather unrelated to the total curvature of the surface).

To make the situation more realistic, one should assume that the billiard ball has a non-zero radius (which would make the rules of reflection more complicated than mathematical billiards usually assume) and then decide what precision of the initial position and velocity of its center of mass is physically meaningful.
To avoid these complications, we prefer to view the billiard ball as a mathematical point whose initial position and velocity are defined perfectly, but consider the sizes $L$ and/or $R$ of the billiard table as physically undefined beyond a certain precision threshold.   

%\begin{figure}[htb]
%\begin{center}
%\epsfig{file=tr}
%\end{center}
%\caption{Detector with a hole}
%\end{figure}

\begin{figure}[htbp]
\centering
\includegraphics[width=0.8\textwidth]{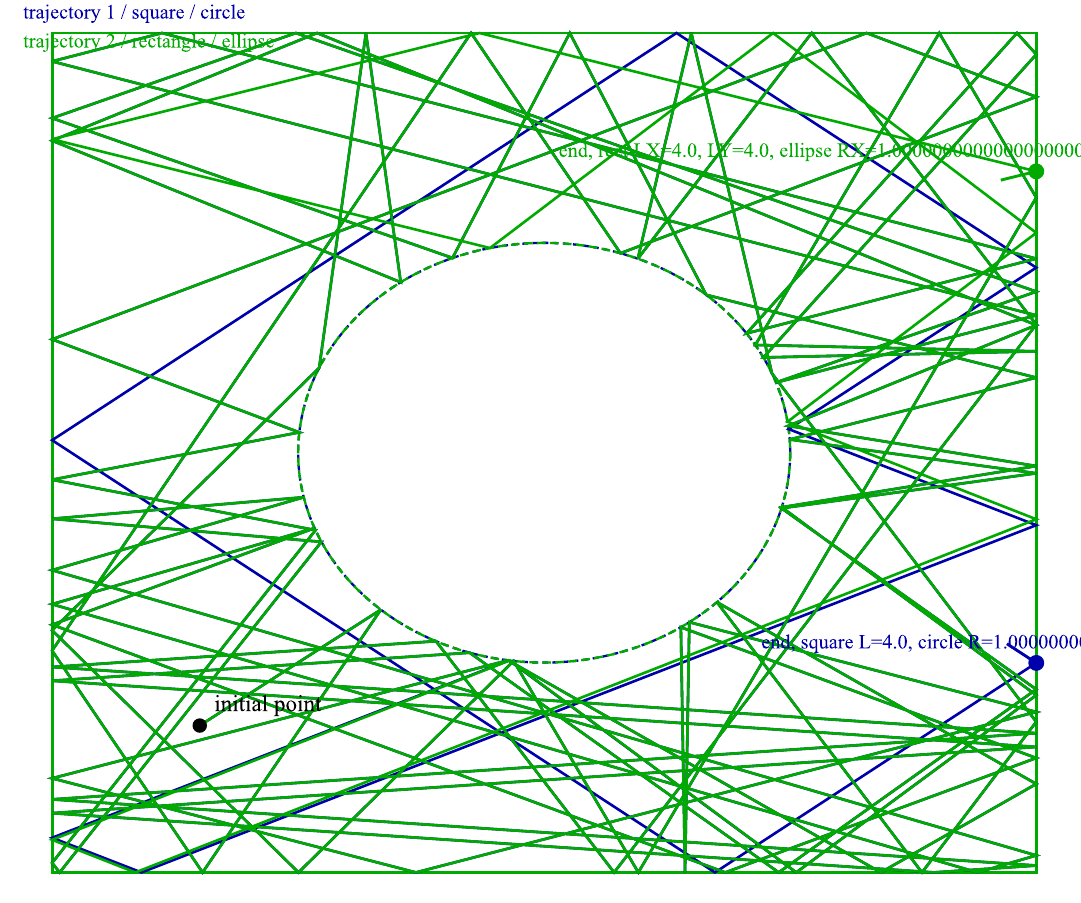}
\caption{$L=4 m$, $R_{green}=1 m$, $R_{blue}=R_{green}+10^{-35} m$}
%\label
\end{figure}

Figure 1 shows in green a billiard trajectory for Sinai's billiard of square size $L=4 m$ and radius $R=1 m$, while a trajectory with the same initial conditions but $R=1+10^{-35}m$, where it is visibly different from the first one, is shown in blue. The separation occurs between 29th and 30th reflection from the disk (and after about 100 reflections overall). With the initial speed $20 m/s$ (the initial velocity vector for this trajectory was $(16, 12)$ with the initial position $(x,y)=(0.6,0.7)$), the separation happened within 9 seconds. Of course, the trajectory does not depend on the actual value of velocity, and so if it is $1 m/s$ then the separation process would take 3 minutes.

Changing the disk into an ellipse of semiaxes which differ by $10^{-35}m$, or similarly altering the size $L$ of the square produce a similar effect.   

\medskip

{\bf 4. Gas dynamics.} Perhaps the most unrealistic aspect of the previous example is the absence of friction and other forms of dissipation: a genuine billiard ball would stop before 30 bounces. However, the system of elastically colliding identical solid balls is considered adequate for modeling a gas of molecules, and it shares some of its features with the 3D version of Lorentz' gas.

Here are the parameters of this system relevant for us. The diameter $d$ of oxygen or nitrogen molecules (repelling each other at close range and hence modeled by solid elastic balls) is considered about $3\cdot 10^{-10}\ m$. From their kinetic energy and mass, the average speed $v$ at room temperature (about $300^{\circ}\ K$) is estimated to be about $500\ m/s$. From the density and the Avogadro number one can estimate typical spacings $L$ between the molecules: a static cubic lattice of such molecules would have the step about $33\cdot 10^{-10}\ m$ (i.e. typical distances are about 10 times the ball's size). Next, the following argument (standard in kinetic theory of gases) allows one to estimate the number $n$ of collisions per second: a ball of radius $d$ moving with speed $v$ sweeps over time $t$ the volume $\pi d^2 vt$ which contains $\pi d^2 v t/ L^3$ molecules. Therefore  
\[  n \sim \frac{\pi d^2}{L^3}v \approx 3\cdot 10^{-2}\frac{500}{3\cdot 10^{-9}} s^{-1}  = 5\cdot 10^9.\]
Consequently, the average distance between collisions\footnote{Taking into account that (unlike Lorentz' lattice) the molecules are not stationary, and their {\em relative} speed matters, the standard formula in the kinetic theory of gases instructs one to divide this answer by $\sqrt{2}$, but this detail doesn't change our conclusions.} $l\approx v/n = 10^{-7} m \approx 30 L$.

Consider now a single event of (non-central) collision of two balls of radius $\approx d/2$, and examine typical consequences of the uncertainty $\delta$ in, say, exact point of contact between the balls (or their radii) at the Planck scale $\delta \sim 10^{-35}\ m$. This uncertainty typically causes the uncertainty
of order $\delta/d$ in the angle of reflection. This change in the angle amplifies the uncertainty of the contact point of the next collision to $\approx \delta l/d$, where $l$ is the distance of the free flight, i.e. typically amplifies the uncertainty by the factor of order $l/d \approx 300$. This leads to the estimate $\log (l/d)$ of the Lyapunov exponent per collision.  Of course, all ingredients in this calculation are random quantities. A shorter free path would result in a smaller amplification factor; alongside the positive local Lyapunov exponent there is the negative one, and so a single collision event may in principle (though unlikely) lead to a decrease of the uncertainty. Consequently, the effective amplification factor is arguably several times smaller than $l/d$.

Let us take a conservative estimate: $L/d \approx 10$ instead of $l/d$. Still,
this means that the initial uncertainty of the Planck scale $10^{-35}\ m$ is amplified to a macroscopic uncertainty in $35$ collisions. In fact even the uncertainty of the size $3\cdot 10^{-10}$, which would require only $25$ collisions of a given molecule already means that it will collide with a different molecule, and so the trajectory of it in the sense of symbolic dynamics is altered irreversibly. However, such subtle distinctions between scales of uncertainty --- initial or final --- don't even matter because of the enormous frequency of collisions. The verdict is that trajectories of individual molecules lose their physical meaning after $10^{-8}\ s$, and of the gas as a whole even several times sooner, e.g. because there is a high certainty that out of $10^{24}$ molecules there are outliers moving significantly faster than average.

In any case, considering a gas of molecules as a deterministic classical mechanical system makes no physical sense past the nanosecond time scale. 

\medskip

As a side remark, let us contrast this conclusion with the celebrated recent mathematical derivation \cite{DHM} of Boltzmann's kinetic equation from the hard sphere model of gas. Here is a heretical thought: What if this and other key principles of statistical physics are justified not by prohibitively monstrous mathematical derivations from the first principles of conservative dynamics, but by augmenting the dynamical models with dissipation accounting for the inherent physical uncertainty of otherwise mathematically precise phase trajectories?   

\medskip

{\bf 5. 3I/ATLAS comet.} According to Kepler's 2-body model, this interstellar visitor of the Solar system arrived along one asymptote of a hyperbolic trajectory and departed along the other one. The theoretical angle $\theta$ between the asymptotes is (as one can calculate) given by the formula
\[ \frac{\theta}{2} = \arctan{\sqrt{e^2-1}} = \arctan\left(\frac{b v^2}{MG}\right),\]
where $e$ is the eccentricity of the hyperbola (a bit above 6 for ATLAS), $MG \approx 1.3 \cdot 10^{20} \ m^3/s^2$ is the product of the Sun's mass $M$ and the gravitational constant $G$, $v\approx 60\ km/s$ is the speed of the comet at infinity (the escape speed for the Sun is about $30\ km/s$), and $b$ is the {\em impact parameter}: the distance by which the asymptote misses the Sun (it is related to the perihelion distance $a$ by $b=a\sqrt{(e+1)/(e-1)}$).  For ATLAS, $a\approx 1.36\ a.u. \approx 2\cdot 10^{11}\ m$, and respectively $b$ is roughly $20\%$ bigger. Thus, in principle the whole event is similar to a reflection of
a material point off a spherical obstruction, except that this time the angle of reflection depends not only on the impact parameter, but also on the speed $v$.

Suppose now that the impact parameter $b$ is determined with a small uncertainty $\Delta b$. Then the uncertainty of $\tan (\theta/2)$ is proportional to it,
\[ \Delta \tan \frac{\theta}{2} = \Delta b \frac{v^2}{MG} = \frac{\Delta b}{b} \tan \frac{\theta}{2}.\]

However, since the derivative of the arctangent function tends to $0$ at infinity, the uncertainty $\Delta \theta$ of the angle becomes very small when $\theta$ is close to $\pi$ (i.e. when the comet passes so far away from the star that the trajectory remains almost straight). So, let's first assume that $\theta$ lies in a ``reasonable'' range so that $\tan (\theta/2) \sim 1$ (about $6.5$ for ATLAS, in which case $\Delta \theta \sim \Delta \tan (\theta/2)$ in the order of magnitude.

Let $L \approx 5\ l.y. \approx 5\cdot 10^{16}\ m$ be the typical distance between stars in the Milky Way. Using the formula $l=L^3/4\pi b^2$ from the gas theory for the average free flight distance, we find that the amplification factor for the uncertainty $\Delta b$ is 
\[ \Lambda = l \frac{\Delta \theta}{\Delta b} \sim \frac{l}{b} \sim \frac{L^3}{b^3} \approx 10^{15}\]
is enormous, and so only a few fly-bys of ATLAS by a star is sufficient to amplify to macroscopic values the initial uncertainty $\Delta b$, however microscopic.

Yet, here is the catch. By the gas theory formulas, the time $t$ over which the number $n$ of fly-bys is about $1$ is estimated as
\[ t \sim \frac{L^3}{4\pi b^2 v} \sim \Lambda \frac{b}{v} \sim 10^{21}\ s, \]
which exceeds the age $\approx 10^{18}\ s$ of the universe by several decimal orders. So, the verdict is that the close fly-by by ATLAS near the Sun is a very unlikely coincidence, the chances of another such fly-by of the same comet by another star are likewise slim, and that of several ones during the comet's lifespan of a few billion years are negligible.

\medskip

The above billiard-inspired simplification of the gravitational problem replaces perpetual interaction of the comet with stars by discrete events when the interaction is turned on only at a close fly-by. Consider, however, a variation of this model where what is meant by fly-bys to be ``close'' is a free parameter. Then we can try to select the value $b$ of this parameter so as to maximize the overall effect of more frequent remote fly-bys ($b$ is large) causing tiny deflection angles $\theta$ with even tinier uncertainties $\Delta \theta$.

To find the value of $b$ optimal for this analysis, let $n=4\pi b^2vt/L^3$ be the number of such fly-bys over a fixed lifespan $t$. Then the average distance between the fly-bys $s=vt/n=L^3/4\pi b^2$, $\Delta \theta /\Delta b \approx 2 GM/ b^2v^2$ (under the assumption that $\tan (\theta/2)$ is large this time), and so
\[ \Lambda(b) = s \frac{\Delta\theta}{\Delta b} \approx \frac{2GML^3}{4\pi v^2b^4} .\]
From this, $b^2$ is proportional to $\Lambda^{-1/2}$ and
\[ n\log \Lambda = C \Lambda^{-1/2} \log \Lambda, \ \ \text{where}\ \ C=\left(\frac{8\pi GM}{L^3}\right)^{1/2}t.\]
The function $\Lambda^{-1/2}\log \Lambda$ achieves maximum at $\Lambda=e^2\approx 7.4$. This corresponds to $C\approx 5\cdot 10^{-15} t$. In other terms:
\begin{align*} b&\approx 6\cdot 10^{15}\ m\approx 4\cdot 10^3\ a.u.\\
  s&\approx 3\cdot 10^{19}\ m \approx 3\cdot 10^3\ l.y.\\
  \tau&:=s/v \approx 5\cdot 10^{14}\ s < 2\cdot 10^7\ \text{years}, \end{align*}
where $\tau$ is the average time between fly-bys, resulting in about $50$ of them per billion years. Consequently the amplification factor $\Lambda^{50}=e^{100}>10^{43}$ per billion years should be sufficient in order to beat the tininess of the initial uncertainty $\Delta b$, provided that it is not excessively tiny though.     

In fact, based on Heisenberg's principle, the initial uncertainty $\Delta b$ can be estimated from below by $\sigma_0:=\sqrt{\tau \hbar/2 m}$. Indeed, if $\Delta b < \sigma_0$, then the uncertainty of the corresponding component of momentum (perpendicular to the trajectory of the comet) $\Delta p > \hbar/2\sigma_0$. This alternative source of uncertainty in the direction of the flight results, after time $\tau$, in the uncertainty of the impact parameter at the next fly-by of about $\tau\Delta p/m > \tau \hbar /2\sigma_0 m =\sigma_0$.\footnote{The value of $\sigma_0$ is known as the {\em standard quantum limit} and can be also obtained as follows. According to the Schr\"odinger equation for a free particle of mass $m$, the initially Gaussian wave packet with $|\psi(0)|^2\sim e^{-q^2/2\sigma(0)^2}$ spreads with time $t$ (remaining Gaussian) with the standard deviation $\sigma(t)$ given by
\[ \sigma^2(t)=\sigma(0)^2+\frac{\hbar^2t^2}{4m^2\sigma(0)^2}.\]
Minimizing over $\sigma(0)$ yields $\sigma(0)=\sqrt{\hbar t/2m}$.}
Thus, either way, values of the impact parameter $b$ make no physical sense below the $\sigma_0$ threshold, and this uncertainty is then dynamically amplified by the Lyapunov factor $\Lambda=e^2$ at subsequent fly-bys (though, with our usual reservations about the random character of this process). 

%According to the Schr\"odinger equation for a free particle of mass $m$, the initially Gaussian wave packet with $|\psi(0)|^2\sim e^{-q^2/2\sigma_0^2}$ spreads with time $t$ (remaining Gaussian) with the standard deviation $\sigma(t)$ given by
%\[ \sigma^2(t)=\sigma_0^2+\frac{\hbar^2t^2}{4m^2\sigma_0^2}.\]
%Minimizing over $\sigma_0$ yields the {\em standard quantum limit} $\sigma_0=\sqrt{\hbar t/2m}$ and $\sigma^2(t)=\hbar t/m$. Note that the optimal quadratic deviation $\sigma_0^2$ merely doubles after time $t$, and remains microscopic for objects of macroscopic mass $m$. As a sanity check, we note that this quadratic deviation of the position measurement after time $t$ is additive with repect to $t$, i.e. doesn't change if intermediate measurements, performed with intervals $t_i$, have statistically independent uncertainties with similarly optimal quadratic deviations $\hbar t_i/m$. Thus, it makes sense to put $t=\tau=s/v$ to obtain

Thus, for our comet (of mass $m$ estimated as $10^{11} - 10^{12}\ kg$) the minimal initial uncertainty of the impact parameter 
\[ \Delta b = \sqrt{\frac{\hbar \tau}{2m}} \approx \sqrt{\frac{10^{-34}\cdot 10^{14}}{10^{12}}} \approx 10^{-16}\ meters.\]
According to our previous estimate this is amplified to the scale $10^{11}$ meters of an astronomical unit after about $30$ remote fly-bys which should take about $600$ million years.

This example suggests that microscopic quantum uncertainties in the position of celestial objects, though remain microscopic under the motion by inertia, can be amplified to macroscopic values during their lifetime under minute gravitational deflections by remote stars.  

\medskip

{\bf 6. Sitnikov's problem.} This is a very degenerate case of the
gravitational 3-body problem. Two identical massive stars revolve around their center of mass $(x,y,z)=(0,0,0)$ along two centrally symmetric elliptic orbits in the $xy$-plane, and the third body is a ``virtually massless'' planet moving along the $z$-axis in the gravitational field of the stars. This system with one degree of freedom is known to exhibit dynamical chaos.

The Hamiltonian of the planet has the form
\[ H(p,q,t)=\frac{p^2}{2}-\frac{2GM}{\sqrt{q^2+r^2(\phi(t))}},\]
where $M$ is the mass of the stars, and $(r,\phi)$ are the polar coordinates of one of the stars with respect to the center of mass and connected by the polar equation of the elliptic orbit:
\[  r (1+E\cos \phi) = \frac{16K^2}{GM} =: l.\]
Here $E$ is the eccentricity of the orbit, and $K$ sectorial velocity, so that the time-periodic function $\phi(t)$ inverses the elliptic integral
\[   \int dt = \frac{1}{2K} \int r^2(\phi) d\phi = \frac{128 K^3}{G^2M^2} \int \frac{d\phi}{(1+E\cos\phi)^2},\]
while the period is found by dividing the area of the ellipse by $K$:
\[ T=\frac{\pi l^2}{(1-E^2)^{3/2}K}=\frac{256 K^3}{G^2M^2 (1-E^2)^{3/2}}.\]
Yet, it is convenient to describe Sitnikov's system in a dimensionless form and using the angle $\phi$ as ``time''. Namely, let $a(\phi)$ be the shorthand for $1+E\cos\phi=l/r(\phi)$. Put $q=lQ$, $p=2KP/l$, and
make the change of time variable
\[\frac{d}{dt}=\frac{d\phi}{dt}\frac{d}{d\phi}=\frac{2Ka^2(\phi)}{l^2}\frac{d}{d\phi}\]
in Sitnikov's Hamiltonian system
\[ \dot{q}=p, \ \ \dot{p}=-\frac{2GMq}{(q^2+r^2(\phi(t))^{3/2}}.\]
This transforms it into
\[ \frac{dQ}{d\phi}=\frac{P}{a^2},\ \ \frac{dP}{d\phi}=-\frac{8aQ}{(1+a^2Q^2)^{3/2}},\]
which is a Hamiltonian system with $2\pi$-periodic Hamiltonian
\[ \H(P,Q,\phi)=\frac{P^2}{2a(\phi)^2}-\frac{8}{a(\phi)\sqrt{1+a(\phi)^2Q^2}}.\]
Here eccentricity $E$ in $a(\phi)=1+E\cos\phi$ is the only parameter.

The next step is to integrate this system numerically, yet using a symplectic integrator, i.e. require that discrete time transformations on the phase plane preserve the symplectic area $dp \wedge dq=2K dP\wedge dQ$. For this, we divide the polar circle into $N$ equal increments $\delta=2\pi/N$, pre-compute the values $a_n:=a(n \delta)$ (depending only on $n \mod N$) of the $\phi$-dependent coefficient $a$, and make each integration step a {\em Chirikov map}:
\[ P_{n+1}=P_n+\delta F(Q_n, a_n), \ Q_{n+1}=Q_n+\delta \frac{P_{n+1}}{a_n^2}=Q_n+\delta \frac{P_n+\delta F(Q_n,a_n)}{a_n^2},\]
where $F(Q,a)=-8aQ/(1+a^2Q^2)^{3/2}$. An easy standard computation shows that its Jacobian is equal to $1$. After $N$ integration steps we obtain a numerical approximation to the {\em Poincar\'e map} of  Sitnikov's system for one revolution of the stars. The plan is to explore the divergence of iterations of this Poincare map applied to close initial positions $(P_0,Q_0)$ on the phase plane.

Let us recall that our ultimate goal is to compare the lifespan of a realistic system with the time interval over which dynamical chaos amplifies to macroscopic values the microscopic intrinsic uncertainties of (say) initial phase points. While Sitnikov's system, whose parameters have clear physical meaning, seems quite suitable for this, we should also realize that in some aspects it remains a toy model. One of them is that it deals with an extremely improbable situation of two identical stars and very special initial conditions of the planet.
As a result, whatever experimental values of Lyapunov exponents one can observe in it, it is unclear how such values would change under a generic perturbation of the system. One can expect a generic system to exhibit more chaotic behavior, but by the same token the long-term behavior of generic trajectories can deviate dramatically from the degenerate ones. Thus, the example at hands is supposed to serve merely as a proof of concept rather than a tool for realistic predictions.

Having said that, we now want to introduce one more aspect that renders our model toy-like. It is related to the discrete nature of numerical simulation. Namely, the approximation of a continuous trajectory of a dynamical system by the number $N$ of discrete steps per period $2\pi$ inevitably introduces an error of order $1/N^2$. (Think of the distance between a circle and the regular $N$-gon inscribed into it.) On the other hand, the scale of intrinsic uncertainties of the initial phase points whose consequences we would like to explore is expressed by a two-digit number of decimal places. Therefore, to make the discretization error insignificant compared to it, we would need $N$, (and consequently the computation time) to be astronomical: on the order of $10^{10}$ or even more. So, our way out of this clash is to take an arbitrarily modest value of $N$, and consider the resulting discrete-time symplectic dynamical system as interesting in its own right: a conceptual analogue of Sitnikov's continuous-time problem,
and not as a faithful numerical approximation to it (which for our purpose it is not). 

The divergence of trajectories of such discrete time Sitnikov-like dynamics is illustrated in Figure 2, where the eccentricity $E=0.85$, $N=36$ iterations per cycle, the initial position $(P,Q)=(-0.5,0.3)$ for one trajectory, and $(-0.5,0.3+10^{-47})$ for the other. The magenta dots show $650$ iterations of the Poincar\'e map for the 2nd trajectory which virtually coincide with those for the 1st one except for the last $50$ or so shown in green.

Assuming that both stars have the mass of the Sun and are 1 a.u. away from their center of mass at the periapsis of their very elongated orbits (apoapsis $\approx 12.33\ a.u.$), one can find that their period of revolution is about $35$ years and so the macroscopic separation of our two trajectories is observed after about $20,000$ years. The initial values $(P,Q)=(-0.5,0.3)$ in dimensionless units corresponds to the meaningful values $q\approx 0.55\ a.u.$ and $\dot{q}\approx -5.5\ km/s$, while $\Delta Q = 10^{-47}$ translates to the Planck scale discrepancy $\Delta q \approx 3\cdot 10^{-36}\ m$ in the initial position. Thus, at least in our toy model, this microscopic uncertainty leads to macroscopic consequences well within the lifespan of a celestial body. Note that the Heisenberg uncertainty $\sigma_0=\sqrt{\hbar t/2m}$ corresponding to the time period $t=20,000$ years and the mass $m \approx 6\cdot 10^{24}\ kg$ of the Earth is about $2\cdot 10^{-24}\ m$ (which in our numerical example led to macroscopic consequences within about $400$ -- $450$ cycles instead of $600$).

\begin{figure}[htbp]
\centering
\includegraphics[width=0.8\textwidth]{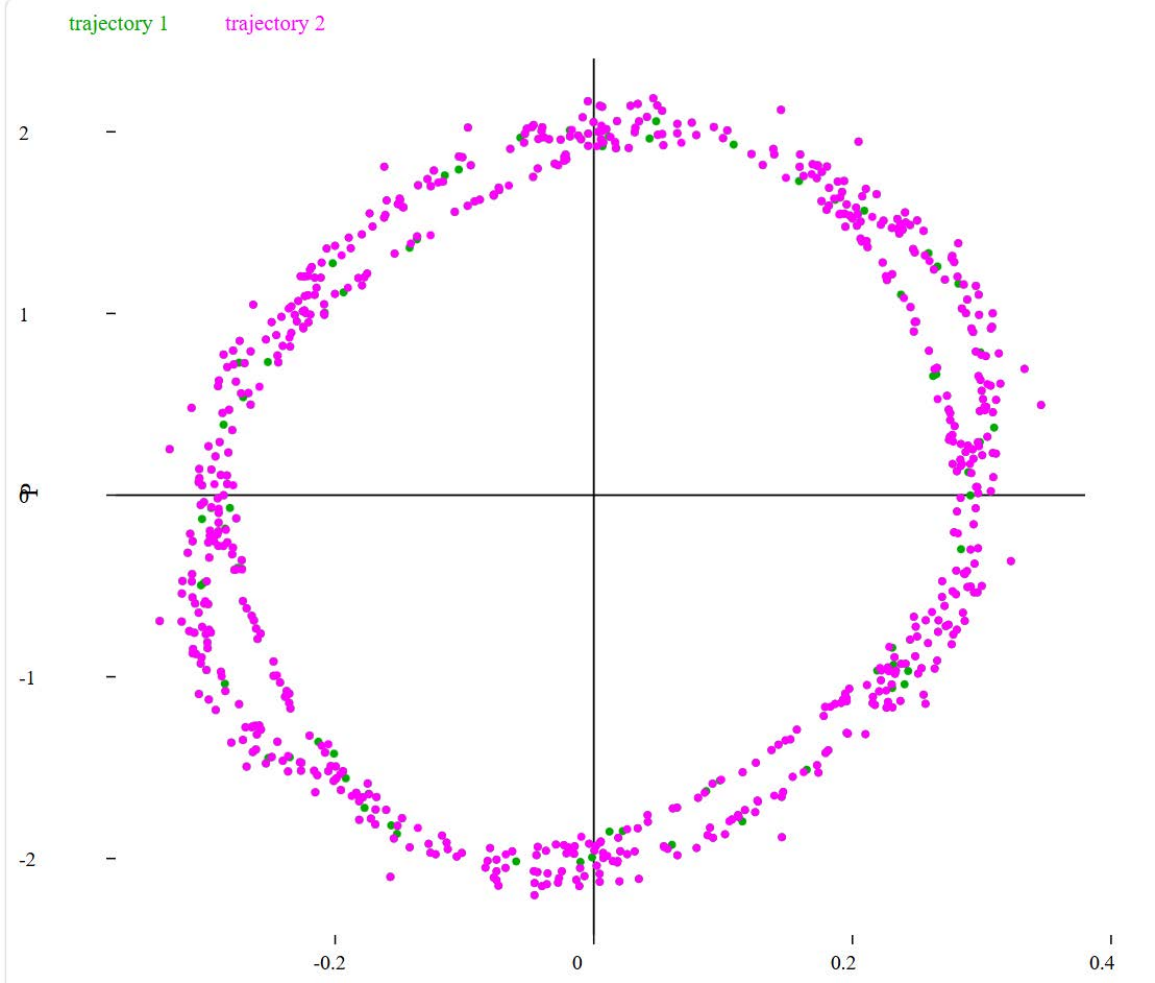}
\caption{$E=0.85, N=36$ \\ $(P,Q)_{green}=(-0.5,0.3)$ \\  $(P,Q)_{magenta}=(P,Q)_{green}+(0,10^{-47})$}
%\label
\end{figure}

{\bf 7. Concluding remarks.} Beside philosophical implications, indeterminism illustrated above can have emotionally significant consequences. An undetected cancerous cell can be carried by the blood stream to an important organ and metastasize or purged out of the body. Is the patient {\em predestined} to respectively die from the disease or survive? I'd expect the answer to be negative: hydrodynamics (of even highly viscous fluids like blood) is a natural source of dynamical chaos, and so it is plausible that the destination of the cell in the body is not merely unpredictable but is inherently uncertain (in the same sense as an outcome of a quantum measurement is genuinely random). It would be interesting to test this hypothesis with some simple models.     

Predestination concerns the future, but Hamiltonian dynamics is time-reversible. Therefore not only {\em retrodiction} is practically unfeasible, but the very certainty of the past is also an illusion. One way to understand this rather mind-bending observation is that the same current state of a system could come from different past states. But there is more to it, and we refer to \cite{GIQ} for further discussion of this in the context of quantum measurement.

I am indebted to Natalie Kozak for her observation about uncertainty of the past, to Svetlana Jitomirskaya for her suggestion to use Sinai's billiards, to ChatGPT for Python coding and numerous brainstorming sessions on all technicalities of this article, and to organizers and participants of the conference ``Billiards and Stars: Geometry and Dynamics'' at CIMAT (Guanajuato, Mexico), dedicated to the 70th anniversary of Richard Montgomery and Sergei Tabachnikov, for the opportunity to present the above material to a competent and friendly audience.

\enddocument